\documentclass[structabstract]{aa}  
\usepackage{graphicx}
\usepackage{txfonts}
\begin{document}
\title{A parsec-scale outflow from the luminous YSO IRAS~17527-2439}


\author{Watson P. Varricatt}

\institute{Joint Astronomy Centre,
   	   660 N. Aohoku Pl., Hilo, HI-96720, USA\\
          \email{w.varricatt@jach.hawaii.edu}
             }

\date{Received Oct 19, 2010; accepted Dec 17, 2010}

 
\abstract
{} 
{We seek to understand the way massive stars form.  The case of a
luminous YSO IRAS~17527-2439 is studied in the infrared.}
{Imaging observations of IRAS~17527-2439 are obtained in the 
near-IR $JHK$ photometric bands and in a narrow-band filter 
centred at the wavelength of the H$_2$ 1-0S(1) line. The 
continuum-subtracted H$_2$ image is used to identify 
outflows. The data obtained in this study are used in 
conjunction with {\it Spitzer}, AKARI, and IRAS data.
The YSO driving the outflow is identified in the 
{{\it Spitzer}} images. The spectral energy distribution 
(SED) of the YSO is studied using available radiative 
transfer models.}
{A parsec-scale bipolar outflow is discovered in our H$_2$ 
line image, which is supported by the detection in the 
archival {{\it Spitzer}} images. The H$_2$ 
image exhibits signs of precession of the main jet and shows 
tentative evidence for a second outflow. These suggest the 
possibility of a companion to the outflow source.  There is 
a strong component of continuum emission in the direction of 
the outflow, which supports the idea that the outflow cavity 
provides a path for radiation to escape, thereby reducing
the radiation pressure on the accreted matter. The 
bulk of the emission observed close to the outflow in the 
WFCAM and {{\it Spitzer}} bands is rotated counter 
clockwise with respect to the outflow traced in H$_2$, 
which may be due to precession. A model fit to the SED of 
the central source tells us that the YSO has a mass of 
12.23\,M$_{\odot}$ and that it is in an early stage of evolution.}
{}


\maketitle
%

\section{Introduction}

Low- and intermediate-mass stars are known to form 
by gravitational collapse and subsequent accretion of 
their parent molecular clouds, and driving collimated 
outflows.  However, the main mechanism leading to the 
formation of massive stars is debated as to whether it 
is either disk accretion similar to that for lower mass stars
(e.g. Yorke \& Sonnhalter \cite{yorke02}) or a merger 
of lower-mass stars (e.g. Bonnell, Bate \& Zinnecker
\cite{bonnell98}). Many of the recent CO line surveys 
show outflows from massive YSOs (e.g. Zhang et al. 
\cite {zhang05}; Beuther et al. \cite{beuther02}).  
Several massive YSO outflows have been observed in 
the near-IR, where the spatial resolution is better 
than that in single-dish CO line observations. A recent 
near-IR imaging survey by Varricatt et al. 
(\cite{varricatt10}) shows that massive stars up to at 
least late-O spectral types form primarily by disk 
accretion.  The case of a luminous YSO taking birth by
accretion is presented in this paper. 

IRAS~17527-2439 (hereafter IRAS~17527) is a luminous 
YSO located in a dark cloud situated close to the 
Galactic plane ($l$ = 4.8273$^{\circ}$, 
$b$ = 0.2297$^{\circ}$) in the 
Ophiuchus region.  It is associated with emission from 
dense gas and dust typical of massive YSOs. Molinari et al. 
(\cite{molinari96}) detected NH$_3$ emission lines from this 
region.  From the radial velocity of the NH$_3$ lines
(V$_{LSR}$=13.3 km~s$^{-1}$), 
they estimated a kinematic distance of 3.23\,kpc, with the 
distance ambiguity resolved. Based on the IRAS colours 
they classified IRAS~17527 as a ``high'' source, 
which is possibly in a UCH{\sc{ii}} phase.  
(Note that at 12\,$\mu$m, the IRAS catalogue gives 
only an upper limit flux for this source). In their 
97.981-GHz CS(2-1) survey of UCH{\sc{ii}} candidates, 
Bronfman, Nyman \& May (\cite{bronfman96}) detected 
IRAS~17527 at V$_{LSR}$=13.5 km~s$^{-1}$, similar to 
the velocity at which Molinari et al. (\cite{molinari96}) 
detected NH$_3$ emission.

Massive star formation is often associated with H$_2$O,
CH$_3$OH and OH maser emission. Palla et al. (\cite{palla91}) 
detected H$_2$O maser emission from IRAS~17527.  From the 
radial velocity of the maser (V$_{peak}$=-1.81~km~s$^{-1}$) 
they estimated a kinematic distance of 17.9\,kpc, which is 
quite different from the distance estimated from radial 
velocity of the  dense gas tracers NH$_3$ and CS. In massive 
YSOs, H$_2$O maser is often considered to be excited in jets 
(e.g. Felli, Palagi \& Tofani \cite{felli92}; Goddi et al. 
\cite{goddi05}).  The blueshift of the wavelength of 
the maser emission in this region with respect to the 
velocity of the dense gas tracers indicates that H$_2$O 
maser near IRAS~17527 also may be excited by a jet.  Hence 
we adopt the distance estimate of 3.23\,kpc (Molinari et al. 
\cite{molinari96}) for the calculations in this paper. 
Surveys by van der Walt, Gaylard \& MacLeod 
(\cite{vanderwalt95}), Walsh et al. (\cite{walsh97}) and 
Slysh et al. (\cite{slysh99}) did not detect any 6.7\,GHz 
Class-II methanol maser from this region.  Edris, Fuller 
\& Cohen (2007) detected faint (0.4\,Jy) OH maser emission 
at V$_{peak}$=11.53\,km~s$^{-1}$, located $\sim$6$\arcmin$ 
NW of IRAS~17527.  Nevertheless,
this offset is less than their beam size. Therefore it remains to 
be investigated at better spatial resolution.  The faintness 
of the detected OH maser is consistent with their observation 
that the OH masers associated with younger sources are weaker 
than the ones associated with H{\sc{ii}} or 
UCH{\sc{ii}} regions. 

The VLA survey of Hughes \& MacLeod (\cite{hughes94}) 
detected 6-cm emission from IRAS~17527 with a peak flux 
of 5.3\,mJy. The source was however rejected as a UCH{\sc{ii}} 
candidate because the radio emission was diffuse.


\section{Observations and data reduction}

\subsection{UKIRT data}
\label{ukirtdata}
Observations were carried out on 2010 Apr. 15 UT using the 
United Kingdom Infrared Telescope (UKIRT) and the Wide 
Field Camera (WFCAM; Casali et al. \cite{casali07}). WFCAM 
has a pixel scale of 0.4$\arcsec$~pix$^{-1}$ and employs 
four 2048$\times$2048 HgCdTe HawaiiII RG arrays.  Each array 
has a field of view of 13.65\,$\arcmin \times$13.65\,$\arcmin$.  
Observations were performed by dithering the object to 9 points 
separated by a few arcseconds and using a 2$\times$2 microstepping.  
Hence the final pixel scale is 0.2$\arcsec$~pixel$^{-1}$. 
Observations were obtained in $J, H$ and $K$ MKO filters and 
in a narrow-band MKO filter centred at the wavelength of the 
H$_2$ (1-0) S1 line at 2.1218\,$\mu$m.   The data were reduced 
by the Cambridge Astronomical Survey Unit (CASU); the archival 
and distribution of the data are carried out by the Wide Field 
Astronomy Unit (WFAU).

\begin{figure}
\centering
\includegraphics[width=8.9cm]{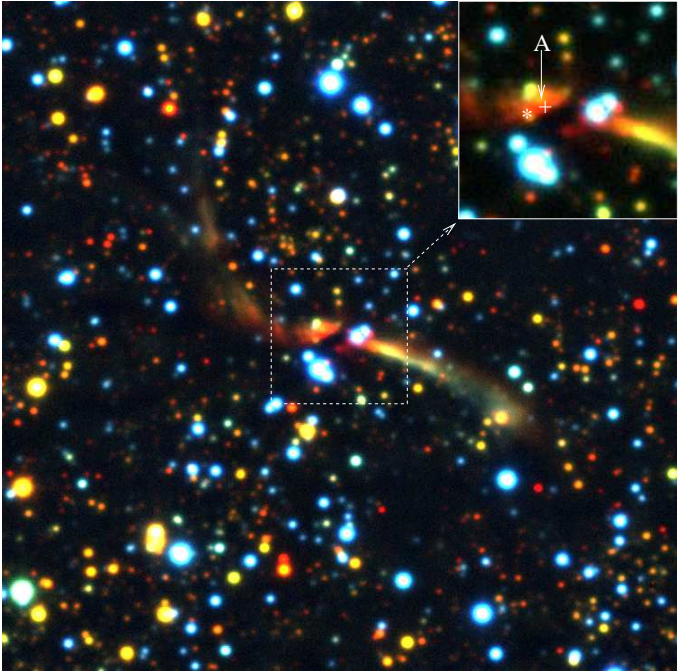}
\caption{WFCAM $JHK$ colour-composite image
($J$-blue, $H$-green, $K$-red) in a
2.5$\arcmin\times$2.5$\arcmin$ field centred
on IRAS~17527. An expanded view of the central
0.5$\arcmin\times$0.5$\arcmin$ field is shown in the inset.
`+' shows the location
of the {\it Spitzer}
source identified and `*' shows the IRAS position.}
\label{wfcamJHK2p5}%
\end{figure}

The sky conditions were not photometric during the observations.
However, since the $K$ magnitude estimated here is
from an average zero point calculated for a set of isolated 
point sources present in all dithered frames around IRAS~17527 
and using their 2MASS magnitudes,  the photometric quality is good. 
Fig. \ref{wfcamJHK2p5} shows a $JHK$ colour composite image 
created from the WFCAM data in a 2.5$\arcmin\times$2.5$\arcmin$
field centred on IRAS~17527. 

\begin{figure*}
\centering
\includegraphics[width=13.5cm]{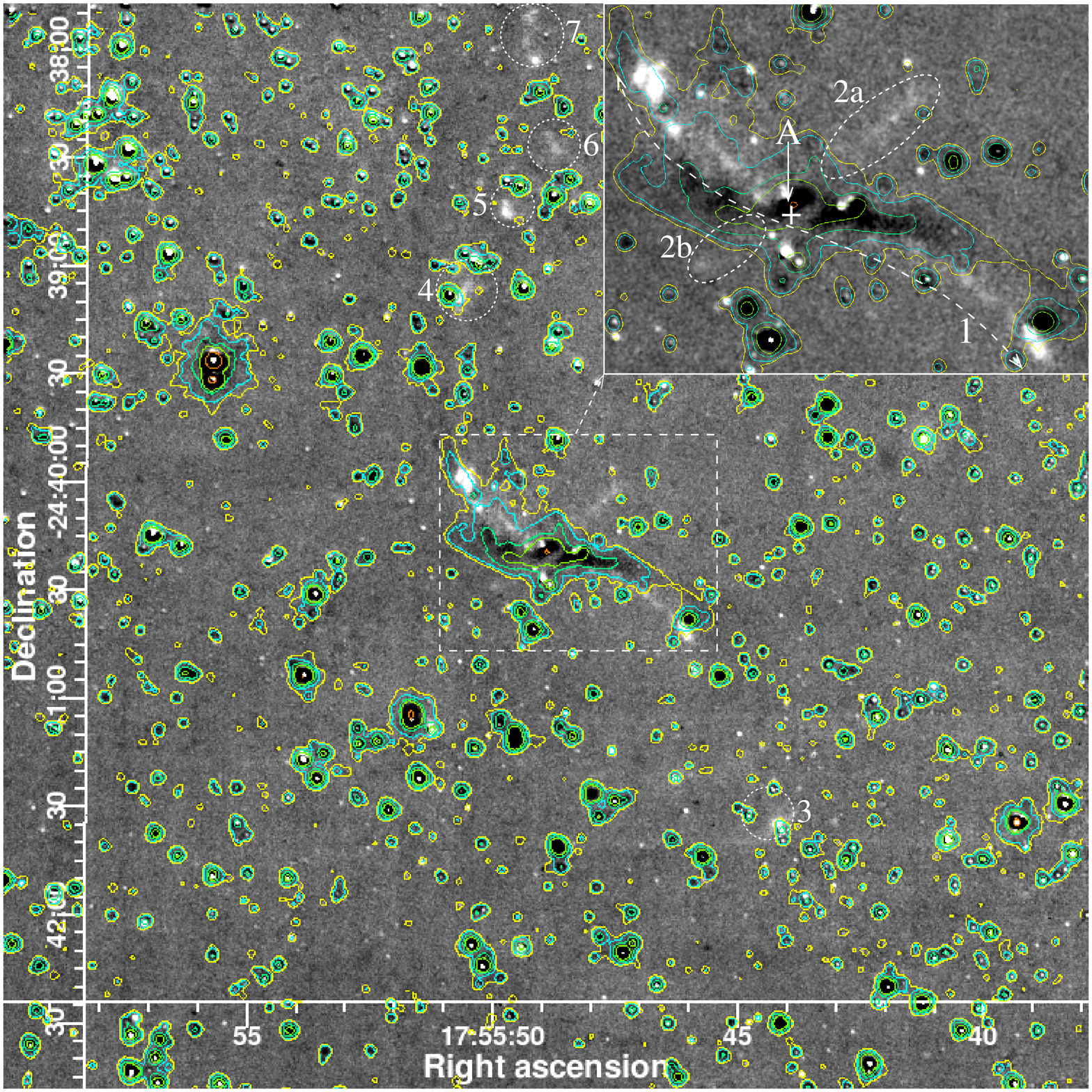}
\caption{The Continuum-subtracted 2.122-$\mu$m image
of IRAS~17527 over a 5$\arcmin$$\times$5$\arcmin$ 
field, smoothed with a 2-pixel FWHM Gaussian to enhance 
the appearance of the faint emission features. `+' shows 
the location of the outflow source derived from the 
{\it Spitzer} 5.8-$\mu$m image.  The contours generated 
from the {\it Spitzer} 4.5-$\mu$m image are overlaid.
An enlarged view of the central region is shown in the
inset.  The dashed line `1' shows the direction of
the main outflow detected here.  The ellipses `2a' and `2b' 
enclose H$_2$ emission features which are probably due
to a second outflow.  Additional H$_2$ emission features
detected are circled and labelled `3--7'.  The image
scale is 0.2$\arcsec$~pix$^{-1}$}.
\label{wfcamH2}
\end{figure*}

The H$_2$ image was continuum-subtracted using the $K$-band 
image.  The average of the ratio of counts, for a few isolated 
point sources, between the $K$ and H$_2$ images was obtained. The 
background-subtracted $K$-band image was scaled by this ratio 
and was then subtracted from the background-subtracted H$_2$
image. In regions where the near-IR colours do not vary widely 
over the sources,  this procedure subtracts out the extended 
emission well, and the point sources get subtracted out well 
when the seeing is similar for $K$ and H$_2$ images. However, 
the $K$/H$_2$ ratio has a dependence on the apparent 
near-IR colours.  For objects like IRAS~17527, which are 
located in dark clouds in the galactic plane, the sources in 
the field are subjected to a wide range of interstellar 
extinction and several sources may have infrared excess due 
to circumstellar emission, resulting in a wide range of near-IR 
colours. Hence, for these regions, even when the seeing is 
stable, a perfect continuum-subtraction cannot be achieved 
for all objects in the field when we use a $K$-band image to 
subtract the continuum from the H$_2$ image.
Fig. \ref{wfcamH2} shows
the continuum-subtracted H$_2$ image.

\subsection {Archival data}

This source was detected well by IRAS at 25 and 60\,$\mu$m. 
At 12\,$\mu$m, the IRAS flux density is only an upper 
limit.  The flux density at 100\,$\mu$m is of poor 
quality and is affected by infrared cirrus.  Hence the 
IRAS 12 and 100-$\mu$m fluxes are not used in our 
analysis. IRAS~17527 was observed in the sky survey 
conducted by AKARI satellite 
(Murakami et al., \cite{murakami07}).  In the mid-IR, 
the Infrared Camera (IRC) on board AKARI detected a 
source at 9 and 18\,$\mu$m. The AKARI Far-Infrared 
Surveyor (FIS) also detected a source in all 
4 bands - at 65, 90, 140 and 160\,$\mu$m.   

The {\it Spitzer} space telescope observed IRAS~17527 as a 
part of the GLIMPSE II survey using the
Infrared Array Camera (IRAC; Fazio et al. \cite{fazio04})
in bands 1--4, centred at 3.6, 4.5,
5.8 and 8.0\,$\mu$m respectively.
{\it Spitzer} also observed this region using the Multiband 
Imaging Photometer for {\it Spitzer} (MIPS; Rieke et al. 
\cite{rieke04}) at 24\,$\mu$m.  The reduced {\it Spitzer} 
images and point source photometry in a 
5$\arcmin\times$5$\arcmin$ field around IRAS~17527 were 
downloaded from the NASA/IPAC Infrared Science Archive.  
Fig. \ref{Spitzer124} shows a colour composite image of 
a 5$\arcmin\times$5$\arcmin$ region constructed from 
the $\it{Spitzer}$ images at 3.6, 4.5 and 8.0\,$\mu$m. 

\section{Results and discussion}

\subsection{The outflow}
\label{outflow}

Our near-IR images 
(Figs. \ref{wfcamJHK2p5}, \ref{wfcamH2}) reveal a 
spectacular outflow oriented NE-SW associated with IRAS~17527.  
A source with large near-IR colours is labelled `A' 
(see $\S$ \ref{source}) in  Fig. \ref{wfcamJHK2p5}.
The continuum-subtracted H$_2$ image (Fig. \ref{wfcamH2}) 
shows the outflow in H$_2$ line emission;  a dashed line 
(labelled `1') is drawn on Fig. \ref{wfcamH2} to guide the 
eyes. The outflow `1' has a length of 48$\arcsec$ in the SW 
and 34$\arcsec$ in the NE if the outflow source is located
at or near `A'.  This length corresponds to 0.75\,pc in the
SW, at a distance of 3.23\,kpc.  The outflow is seen in H$_2$ 
to be bent in an `S'-shaped fashion, suggesting a precession 
of the jet.  The direction of the outflow measured from 
the H$_2$ images is $\sim$68$^\circ$ North of East at the base.   
The detection of a well defined outflow suggests that 
the YSO associated with IRAS~17527 is taking birth through 
disk accretion.  There is also tentative detection of two 
additional H$_2$ emission features close to IRAS~17527, 
which are enclosed in ellipses on Fig. \ref{wfcamH2} and 
labelled `2a' and 
`2b'; this needs to be verified by deeper H$_2$ imaging. 
If a second bipolar outflow is producing `2a' and `2b', it
may be at an angle of 209$^\circ$. Thus, the possibility 
of a companion needs to be explored through deeper H$_2$ imaging 
and through imaging in the near- and mid-IR at higher angular 
resolution.  The model fit to the SED (see $\S$ \ref{fitting}) 
gives an angle of inclination of the disk axis with
respect to the line of sight to be 18$^{\circ}$, 
which implies a similar line-of-sight inclination 
for the 
outflow. The comparable brightness for the two outflow lobes 
of `1' seen in H$_2$ suggests a higher inclination 
(even though it could be argued that this is mainly due to 
non-uniform distribution of exciting material rather than 
higher inclination). 
The outflows `1' and `2a+2b' are assigned 
names MHO~2147 and MHO~2148 respectively in the catalog of
Molecular Hydrogen emission-line Objects{\footnote 
{http://www.jach.hawaii.edu/UKIRT/MHCat/} (Davis et al. \cite{davis10}).
Some additional line emission features detected in the
continuum-subtracted H$_2$ image are circled and
labelled `3--7' in Fig. \ref{wfcamH2}.  It is not clear
if these are produced by outflow `1' or by different
outflows.  

\begin{figure}
\centering
\includegraphics[width=8.9cm]{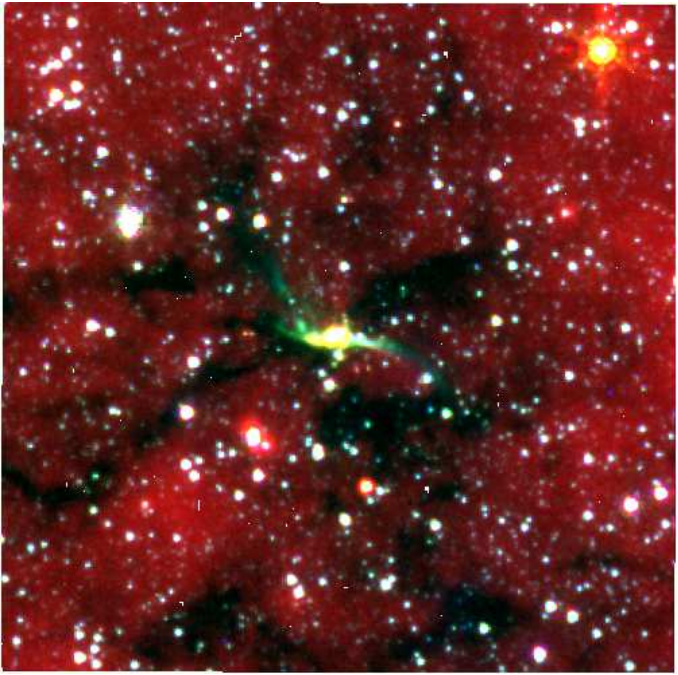}
\caption{A colour-composite image produced from {\it Spitzer}-IRAC
data adopting blue, green and red, respectively for the 3.6,
4.5 and 8.0-$\mu$m images. The image covers a
5$\arcmin \times$5$\arcmin$ field  centred on IRAS~17527.}
\label{Spitzer124}%
\end{figure}

There is a significant amount of emission along the outflow
in our $JHK$ and unsubtracted H$_2$ images and in the 
{\it Spitzer}-IRAC images.  A major portion of the 
emission seen along the outflow in the H$_2$ image is subtracted 
out upon continuum-subtraction and Fig. \ref{wfcamH2} shows 
only a chain of line emission knots.  This suggests the presence 
of a large amount of continuum emission in the direction the 
outflow. Continuum emission close to the outflow is likely to 
be due to radiation from the YSO escaping through the outflow 
cavity.  It has been proposed before that in massive star 
formation, the optically thin cavity carved out by the bipolar 
outflow may provide a path for radiation from the central 
source to escape, which reduces the radiation 
pressure on the accreted matter and aids the growth of a 
massive star. (Krumholz, McKee \& Tan 
\cite{krumholz05}).

Fig. \ref{wfcamH2} exhibits negative residuals along 
the bright regions of emission in the near-IR images, seen 
as dark lanes close to the outflow traced by the H$_2$ 
knots labelled `1'.  The negative residuals 
close to the outflow are likely to be caused by large near-IR 
colours arising from extinction or excess or both, as 
discussed in $\S$ \ref{ukirtdata}.  The contours 
generated from the {\it Spitzer}-IRAC 4.5-$\mu$m image 
are over-plotted in  Fig. \ref{wfcamH2}.  Away from the 
outflow centre, the contours of the faint regions of 
the 4.5-$\mu$m emission trace the outflow detected in H$_2$ 
in the NE, but are rotated counterclockwise with respect to 
the H$_2$ knots in the SW.  Close to the base of the outflow, 
the contours of the bright regions of the 4.5-$\mu$m
emission are nearly EW.   The 4.5-$\mu$m contours trace 
the bright regions of the $K$-band emission reasonably well; 
close to the base of the outflow, the contours of the bright
regions of the 4.5-$\mu$m emission have a slight counterclockwise 
rotation with respect to the bright regions of the emission in 
$K$ in the east.  In general, the bright regions of emission 
seen in the near-IR and {\it Spitzer} images, close to the 
outflow, are rotated counter-clockwise with respect to 
the direction of the outflow implied by the presumably 
shock-excited H$_2$ emission knots in `1'.  This rotation, 
again, is likely to be due to the precession of the jet, but
needs to be thoroughly investigated.  The brightest region of 
this emission in the near IR towards the SW lobe of the 
outflow is seen throughout the 1--8\,$\mu$m wavelength range.  
It should be noted that transitions of H$_2$ are present 
throughout the 1--8\,$\mu$m wavelength range discussed here 
(Smith \cite{smith95}; Neufeld et al. \cite{neufeld09}). 
Spectroscopy in the near- and mid-IR will enable us to 
understand if the emission seen in in these bands
close to the outflow `1' are line- or 
continuum-dominated. If continuum-dominated, 
the emission, especially in short wavelengths like $J$, 
will imply radiation escaping from the hot central source.

\subsection{The source driving the outflow}
\label{source}

The bright objects and a significant fraction of
the reddened objects in the WFCAM images are detected in 
the IRAC images, as can be seen from the IRAC 4.5-$\mu$m 
contours around the point sources in the continuum-subtracted
H$_2$ image (Fig. \ref{wfcamH2}).  The most prominent among
these is a bright infrared source detected near the 
centre of the outflow lobes in the IRAC and MIPS images. 
This source is saturated in the 24-$\mu$m MIPS image.
Its coordinates derived from the IRAC 8.0-$\mu$m image
are $\alpha$=17:55:48.91, $\delta$=-24:40:19.9\footnote{all 
coordinates given in this paper are in J2000}; it
is only at a separation of 2$\arcsec$ from the IRAS 
position. The IRAC catalog does not list the magnitudes 
for this source in the 3.6 and 4.5-$\mu$m bands.  At 
these wavelengths, the source appears elongated 
in the {\it Spitzer} images and the emission from it is 
blended with the emission from the outflow.  For the 
purpose of fitting the SED, we derived the 3.6 and 
4.5-$\mu$m magnitudes of the outflow source from the 
{\it Spitzer} images. Relative photometry was performed 
by adopting an aperture of 8-pixel (4.8$\arcsec$)
diameter.  For each band, an average zero point was
derived
from a few well detected and isolated point sources
(which have IRAC magnitudes available)
spread over the image.  We derived a magnitude of 
9.33$\pm$0.03 in the 3.6-$\mu$m band and 7.45$\pm$0.05 in the 
4.5-$\mu$m band.  The errors given are the formal errors in the 
zero points. Even at this aperture, the emission from the 
outflows will be contributing to the observed fluxes. For the 
SED fitting, we adopted an error of 0.18 mag. 

We identify the source detected in the {\it Spitzer} data 
as the YSO driving the outflow `1'. A deeply embedded object 
seen only in $K$ and H$_2$, located near the centre of the 
outflow, is labelled `A'  ($\alpha$=17:55:48.95, 
$\delta$=-24:40:18.31) in Fig. \ref{wfcamJHK2p5}. `A' 
appears slightly extended in $K$ and is surrounded by 
nebulosity.  Its coordinates are offset 1.65$\arcsec$ NE from 
that of the {\it Spitzer} source measured at 8\,$\mu$m.  
Photometry within an aperture of 2$\arcsec$ diameter gives 
a $K$ magnitue of 12.82$\pm$0.04 for `A'.
It remains to be investigated through high angular resolution 
imaging at near- and mid-IR wavelengths to understand if `A' 
is the near-IR counterpart of the outflow source.
The $K$ magnitude of `A'  
is close to what is implied by the fit to the SED (Fig. \ref{17527_sed}).
However, ``A'' has a positional offset from the {\it Spitzer}
source, which, even though small, is significant, and the 
SED fitting gives an inclination of 18$^{\circ}$ with 
respect to the line of sight for the disk axis. 
From the appearance of the outflow `1', the angle appears to 
be higher ($\S$ \ref{outflow}).  A higher inclination of the 
disk will pose a higher circumstellar extinction for the 
outflow source, making it too faint to detect in the 
near-IR. In addition, the the magnitude of `A' estimated 
here is likely to be 
contaminated by nebulosity.  With these considerations, we 
are treating the $K$ magnitude of `A' as only an upper 
limit for modelling the SED of the YSO ($\S$ \ref{fitting}).


\begin{figure}
\centering
\includegraphics[width=8.9cm]{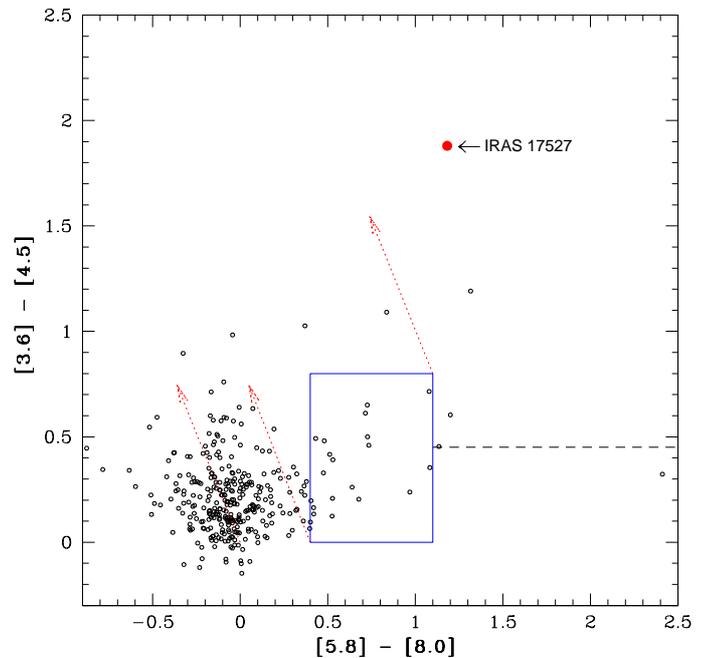}
\caption{The IRAC colour-colour diagram.  The open 
(black) circles show the colours of objects detected 
in all four bands in a 5$\arcmin\times$5$\arcmin$
field centred on IRAS~17527. The filled (red) circle 
shows the outflow source.  The blue rectangle shows
the location of Class II sources.  The dashed 
horizontal line shows the boundary between Class I/II 
objects (below) and Class I objects (above).  
The dotted red arrows show reddening vectors for
A$_V$=45.
}
\label{IRAC_col}%
\end{figure}

IRAS~17527 is detected well in all mid- and far-IR bands of
AKARI satellite.  The coordinates given in the AKARI - IRC point 
source catalog is only 1.2$\arcsec$ NW 
of the {\it Spitzer} position. At far-IR wavelengths, 
AKARI - FIS detected a bright object 
(AKARI-FIS-V1 1755489-244001; $\alpha$=17:55:48.89  
$\delta$=-24:40:0.82) in all 4 bands.
The coordinates of the FIS detection is offset 
18.75$\arcsec$ towards the north of the source detected by IRC. 
This offset is  more than the positional accuracy quoted by 
the AKARI point source catalogue.  However, since it is the 
only far-IR source in the region, and the coordinates of the 
infrared source detected in this region by {\it Spitzer} and 
IRAS agree well, we ascribe the FIS detection to IRAS~17527.

The IRAC~([5.8]-[8.0],[3.6]-[4.5]) colour-colour diagram
can be used to roughly identify YSOs of different evolutionary
stages. Fig. \ref{IRAC_col} shows the IRAC colours of objects, 
within a 5$\arcmin \times$5$\arcmin$ field around IRAS~17527. 
The large concentration of sources around 
([5.8]-[8.0],[3.6]-[4.5])=0 are foreground and background 
stars and diskless pre-main-sequence (Class~III) objects. 
The dashed horizontal line shows the boundary  between 
Class I/II 
(below) and Class~I objects (above).  The rectangle
(0$<$([3.6]-[4.5])$<$0.8 and 0.4$<$([5.8]-[8.0]$<$1.1) shows 
the location of Class II objects. The sources with 
([3.6]-[4.5])$>$0.8 and ([5.8]-[8.0])$>$1.1 are likely to 
be Class~I objects, which are protostars with infalling 
envelopes (Allen et al. \cite{allen04}; Megeath et al. 
\cite{megeath04}; Qiu et al. \cite{qiu08}).  The 
dotted red arrows in Fig. \ref{IRAC_col} are reddening 
vectors for A$_V$=45 calculated from the extinction law 
given in Mathis (\cite{mathis90}).  The location of 
IRAS~17527 in Fig.\ref{IRAC_col} suggests that it is 
a highly reddened Class~I object. As seen from the SED 
modelling in Fig. \ref{17527_sed}, the emission from 
the envelope contributes significantly to the luminosity,
which is expected for Class~I objects.

\subsubsection{The spectral energy distribution}
\label{fitting}
The $K$ magnitude measured from the WFCAM data, magnitudes 
in the {\it Spitzer} bands 1--4, IRAS 25 and 60-$\mu$m data 
and the AKARI data from 9 to 160\,$\mu$m were used to 
construct the SED of the source driving the outflow. Colour 
corrections were applied to the IRAS and AKARI data using 
the correction factors given in their respective point source 
catalogs (Beichman et al. \cite{beichman88};  Kataza et al. 
\cite{kataza10}; Yamamura et al. \cite{yamamura10}).  The 
SED was modelled using the on-line SED fitting tool of 
Robitaille et al. (\cite{robitaille07}), which uses a grid of
2D radiative transfer models presented in Robitaille et al.
(\cite{robitaille06}), developed by Whitney et al. 
(\cite{whitney03a,whitney03b}; etc.).  The grid consists of 20,000 
YSO models with SEDs covering a wide range of stellar 
masses (0.1--50\,M$_\odot$) and evolutionary stages (from 
the early envelope infall stage to the late disk-only 
stage), each at 10 different viewing angles.

\begin{table}
\caption{Results from SED fitting}             
\label{tab:results}      
\centering                      
\begin{tabular}{ll}        
\hline\hline					\\[-2mm]
Parameter                       &Best fit values$^{\mathrm{a}}$		\\
\hline              				\\[-2mm]         
Stellar mass (M$_{\odot}$)      &12.23 (+1.4, -0.97)			\\
Stellar age (yr)                &2.08 ($\pm$1)$\times$10$^{4}$		\\
Stellar radius (R$_{\odot})$	&39.94 (+54, -12.9)			\\
Stellar temperature (K)	&8093 (+2000, -2600)				\\
Disk mass (M$_{\odot}$)         &9.79 (+29, -7.3)$\times$10$^{-2}$	\\
Disk accretion rate (M$_{\odot}$yr$^{-1}$)  &4.16 (+139, -3.2  )$\times$10$^{-6}$	\\
Disk/envelope inner radius (AU) &5.91 (+7.9, -0.83)			\\
Disk outer radius (AU)		&28.6 (+69, -12)			\\
Envelope mass (M$_{\odot}$)     &1.56 (+1, -0.67)$\times$10$^{3}$    	\\
Envelope accretion rate (M$_{\odot}$yr$^{-1}$)  &2.42 (+1.9, -0.84)$\times$10$^{-3}$ \\
Envelope outer radius (AU)         &1.00 (+0, -0.079)$\times$10$^{5}$	\\
Total Luminosity (L$_{\odot}$)  &6.17 (+2, -1.5)$\times$10$^{3}$	\\
Angle of inclination of the disk axis ($^{\circ}$)	&18		\\

\hline 
\end{tabular}
\begin{list}{}{}
\item[$^{\mathrm{a}}$] The values given in parenthesis are the rms of 
the differences, in parameter values of models with 
($\chi^2-\chi^2_{best fit}$) per data point $<$ 3, above and below those 
of the best fit model, estimated with respect to the 
parameter values of the best fit model.  A distance of 3.23\,kpc 
is adopted for IRAS~17527.
\end{list}
\end{table}

Fig. \ref{17527_sed} shows the SED of IRAS~17527 
and the model fit. Table \ref{tab:results} shows the 
parameters of the best fit model.  The foreground visual 
extinction A$_V$ is estimated to be 45.43 mag.
The fitting
gives a stellar mass of 12.23\,M$_{\odot}$ and a 
total luminosity of 6.17$\times$10$^{3}$\,L$_{\odot}$.
The mass of the YSO and the systemic luminosity are 
reasonably well determined, so are the envelope parameters.
However, the radius of the YSO, the disk
accretion rate and the disk mass and radii have
large errors. The near- and mid-IR fluxes have a significant
influence on the reliable determination of these parameters.
Contamination of the source magnitudes at these wavelengths
due to the contribution from the outflow and from a possible 
companion may be the cause for the poor estimate of
these parameters.  Hence, 
photometry at high sensitivity and spatial resolution in 
the 2--20-$\mu$m wavelength range is warranted for a more 
accurate determination of the stellar and disk parameters.
Photometry at sub-mm wavelengths also is required
for this source to better constrain the model fit. 

\begin{figure}
\centering
\includegraphics[width=8.9cm]{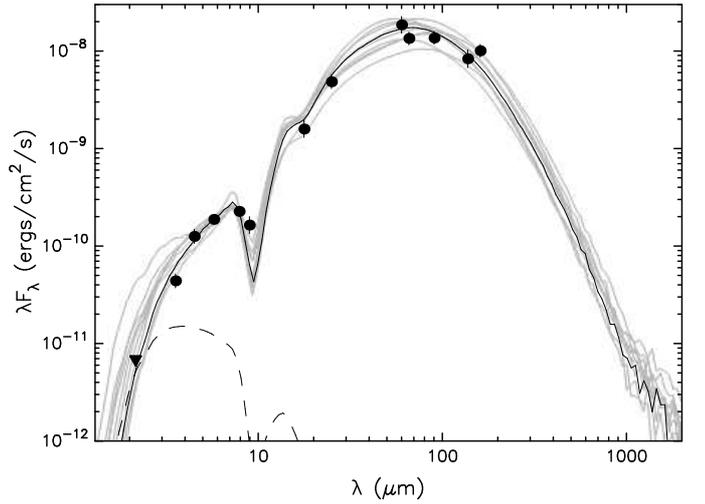}
\caption{The filled circles show the data from {\it Spiter}-IRAC
at 3.6, 4.5, 5.8 and 8\,$\mu$m, IRAS at 25 and 60\,$\mu$m, and AKARI 
at 9, 18, 65, 90, 140 and 160\,$\mu$m. The downward 
directed triangle shows the $K$-band data, which is
treated as an upper limit only. The continuous line
shows the best fit model and the gray lines show subsequent good 
fits for ($\chi^2-\chi^2_{best fit}$) per data point $<$ 3.
The dashed line corresponds to the stellar photosphere for the 
central source of the best fitting model, as it would look 
in the absence of circumstellar dust (but including interstellar 
extinction).}
\label{17527_sed}%
\end{figure}

\section{Conclusions}

\begin{enumerate}
\item IRAS~17527 appears to be a luminous YSO taking 
birth through disk accretion. A well-collimated 
parsec-scale outflow is discovered in IRAS~17527 in H$_2$
line emission. 
\item The outflow  is seen 
to be bent in an `S'-shaped fashion, suggesting a
precession of the jet. There is a tentative detection 
of a second outflow in the H$_2$ image. The 
possibility of more than one YSO in IRAS~17527 needs to be 
explored. 
\item The H$_2$ image exhibits a
strong continuum component in the emission along the 
outflow, which may be caused by radiation escaping through the outflow 
cavity. This, in turn, reduces the 
radiation pressure on the accreted matter and aids growth of the 
central source through accretion.
\item The bulk of the emission in the direction of the outflow,
observed in $K$ and the {\it Spitzer} bands, is rotated 
counter-clockwise with respect to the direction of the outflow 
traced by the H$_2$ line emission knots.  It is probably a
result of the precession of the jet.  Spectroscopy in the
the near-IR through {\it Spitzer}-IRAC bands is required 
for a proper understanding of this.
\item The model fit to the SED shows that the central 
source is probably a Class-I protostar; this is supported 
by its location in the {\it Spitzer}-IRAC 
colour-colour diagram.  The YSO has a mass of 
$\sim$12.23\,M$_{\odot}$ and a total luminosity 
of $\sim$6.17$\times$10$^3$\,L$_{\odot}$. 
\item The disk parameters and the disk accretion
ratio are poorly determined by the SED fitting.  This may 
be caused by the contribution to the source magnitudes from 
emission from the outflow and from a possible companion.
Observations at high angular 
resolution are required in the 2-20\,$\mu$m wavelength 
range for a more accurate determination of the source
magnitudes at these wavelengths.
\end{enumerate}

\begin{acknowledgements}
The UKIRT is operated by the Joint Astronomy Centre on behalf 
of the Science and Technology Facilities Council (STFC) of the 
UK.  The UKIRT data presented in this paper are obtained during 
the UKIDSS back up time. I thank the Cambridge Astronomical 
Survey Unit (CASU) for processing the WFCAM data, and the WFCAM 
Science Archive (WSA) for making the data available.  This
work makes use of data obtained with AKARI, a JAXA project with
the participation of ESA, and IRAS data downloaded from the
SIMBAD database operated by CDS, Strasbourg, France.
The {\it Spitzer} archival data are downloaded from NASA/ IPAC 
Infrared Science Archive, which is operated by the Jet Propulsion 
Laboratory, Caltec, under contract with NASA. 
I also thank the anonymous referee and the editor Malcolm
Walmsley for their comments and suggestions
which have improved the paper.
\end{acknowledgements}

\end{document}